\documentclass[copyright,creativecommons]{eptcs}
\providecommand{\event}{QAPL 2014} % Name of the event you are submitting to
\usepackage{breakurl}             % Not needed if you use pdflatex only.
\usepackage{fancyhdr}
\usepackage{amsmath,amsthm,amssymb}
\usepackage{graphicx}
\usepackage{hyperref}
\usepackage{listings}
\usepackage{subfigure}
\usepackage{multirow}
\usepackage{array}
\usepackage{float}
\usepackage{paralist}

\def\S{\mbox{\large $\rhd\;\!\!\!\!\!\!\!\lhd$}}
\def\rmdef{\stackrel{\mbox{\em {\tiny def}}}{=}}

\newcommand{\sync}[1]{\raisebox{-1.0ex}{$\;\stackrel{\S}{\scriptscriptstyle #1}\,$}}

\newcommand{\pc}{{\,:\,}}
\newcommand{\Sys}{\mathit{Sys}}
\newcommand{\ZebraNetCtrl}{\mathit{ZebraNetCtrl}}

\newcommand{\IncA}{\mathit{IncA}}

\newcommand{\rmdefl}{\smash{\raisebox{-2.0pt}[0pt][0pt]{$\,\rmdef\,$}}}

\newcommand{\ev}[1]{\text{\underbar{#1}}}

\newcommand{\ec}{\mathop{\mathit{ec}}}

\newcommand{\const}{\mathop{\mathit{const}}}

\newcommand{\Con}{\mathop{\mathit{Con}}}

\newcommand{\icbisim}{\mathop{\raise-1pt\hbox{$\stackrel{\raise-1pt\hbox{\small
$\leftrightarrow$}}{=}$}}}
\newcommand{\bmbisim}{\mathop{\stackrel{\raise-10pt\hbox{\small
$\leftrightarrow$}}{\hbox{\rule{8pt}{0.4pt}}}}}

\title{Patch-based Hybrid Modelling of Spatially Distributed Systems by Using Stochastic HYPE - ZebraNet as an Example}
\author{Cheng Feng
\institute{Laboratory for Foundations of Computer Science\\
University of Edinburg\\
Scotland, UK}
\email{s1109873@sms.ed.ac.uk}
}
\def\titlerunning{Patch-based Hybrid Modelling of Spatially Distributed Systems by Using Stochastic HYPE}
\def\authorrunning{Cheng Feng}
\begin{document}
\maketitle

\begin{abstract}
Individual-based hybrid modelling of spatially distributed systems is usually expensive. Here, we consider a hybrid system in which mobile agents spread over the space and interact with each other when in close proximity. An individual-based model for this system needs to capture the spatial attributes of every agent and monitor the interaction between each pair of them. As a result, the cost of simulating this model grows exponentially as the number of agents increases. For this reason, a patch-based model with more abstraction but better scalability is advantageous. In a patch-based model, instead of representing each agent separately, we model the agents in a patch as an aggregation. This property significantly enhances the scalability of the model. In this paper, we convert an individual-based model for a spatially distributed network system for wild-life monitoring, ZebraNet, to a patch-based stochastic HYPE model with accurate performance evaluation. We show the ease and expressiveness of stochastic HYPE for patch-based modelling of hybrid systems. Moreover, a mean-field analytical model is proposed as the fluid flow approximation of the stochastic HYPE model, which can be used to investigate the average behaviour of the modelled system over an infinite number of simulation runs of the stochastic HYPE model.
\end{abstract}

\section{Introduction}
Spatially distributed systems are encountered in a variety of natural and engineering scenarios. Individual-based modelling of such systems suffers from its low level scalability as the cost of analysing these models depends on the number of entities in the system. Thus, patch-based modelling in which entities are grouped according to their physical positions can be superior. More specifically, in a patch-based model, the space is divided into discrete patches (locations, cells, islands, etc.),  the entities within a patch are assumed to share similar attributes. As a result, there is no need to capture the attributes of every individual. The cost of analysing the model is dependent on the number of patches instead of entities in the system. This property significantly improves the scalability of the model.

In this paper, we present a patch-based stochastic HYPE model for a spatially distributed system which is originally analysed by an individual-based simulation program. The modelling language, stochastic HYPE \cite{galpin2009hype, bortolussi57hype}, is a process algebra, meaning that it is equipped with a formal interpretation in terms of an underlying mathematical model and equivalence relations. We will show the ease and expressiveness of stochastic HYPE in patch-based hybrid modelling through the introduction of the model. 

The example system which we consider is ZebraNet \cite{zebraNet}, a sensor network deployed in central Kenya to collect data on zebras for biological research. In ZebraNet, zebras are fitted with collars which collect and transmit data about zebras' movements, temperatures, etc. Zebras are naturally distributed over a large area. Whenever two zebras are in close proximity, they exchange all their stored data (relating to themselves and other zebras) with each other (using a flooding protocol). Periodically, a mobile base station circulates to collect data from zebras for further biological research. The reason why we choose ZebraNet as our case study is because there is a well-described existing individual-based simulation program, ZNetSim \cite{zebraNet} (written in C) for ZebraNet. In ZNetSim, each zebra's behaviour is simulated explicitly and separately. This approach is not suitable for analysing the system when the number of zebras in the system is large. Thus, we build a patch-based stochastic HYPE model for ZebraNet in which the map is represented by patches and the zebras' movement is captured in terms of going from one patch to another. We assume that all zebras within a patch share data and therefore we can think in terms of the data of that patch. We use the \emph{age of data} to denote the length of time since the last time the mobile base station or other patches received fresh data from a particular patch and use it to compute the success rate of data delivery to the mobile base station. The patch-based HYPE model is suitable for analysing the system with an arbitrary number of zebras, and can give accurate performance evaluation, which is validated by comparing our simulation results with ZNetSim.

In addition, a mean-field analytical model is presented to describe the stochastic HYPE model for ZebraNet by a set of ordinary differential equations (ODEs). In the mean-field model, we treat the evolution of all the variables in the stochastic HYPE model as fluid flows. The mean-field model reveals the average value of the variables in the stochastic HYPE model over an infinite number of simulation runs. It can give a computationally efficient way for evaluating the average behaviour of the underlying modelled system.

% we present a formal modelling approach which is intended to assist in the design and verification of both functional and non-functional properties of hybrid systems, and demonstrate its use on a patch-based model for a %spatially distributed system. 

%Unlike the existing process algebras for hybrid systems which model the continuous variables in the hybrid systems monolithically, HYPE offers a compositional approach for modelling hybrid systems. In HYPE, the various %continuous and discrete behaviour of the system can be modelled separately, this property significantly improves the ease and flexibility of HYPE in modelling hybrid systems. 

The paper is structured as follows. After introducing the background of ZebraNet and HYPE in more detail, we present the patch-based stochastic HYPE model for ZebraNet. We evaluate the model and compare our experiment result with the original simulation program ZNetSim \cite{zebraNet} in Section \ref{evalu}. Section \ref{meanfield} presents the mean-field analytical model. Finally, Sections \ref{relatedworks} and \ref{conclusions} discuss related work, future research and draw final conclusions.

\section{Background}
\subsection{ZebraNet}
ZebraNet is an opportunistic sensor network that is deployed in central Kenya to collect data on zebras for biological research. In the original simulation program ZNetSim, 50 zebras wearing special collars, as well as 10 water sources, are randomly placed across a $20km\times 20km$ map. Zebras have three movement patterns, which are grazing, grazing-walking and fast-moving. Different movement patterns mean different moving speed and frequency of turning angles. Zebras also get thirsty once each day. When they get thirsty, they head to their nearest water source directly using constant speed. After they reach the water source, they move randomly on the map as usual\footnote{A detailed introduction of zebras' movement patterns can be found in \cite{zebraNet}}. The collars collect data on zebras every three minutes. Moreover, zebras will flood their stored data for themselves and others to all neighbours when they are discovered within the 100 meters peer discovery range. A mobile base station, which follows a rectangular route, will periodically collect data from zebras. More specifically, the zebras transmit all their stored data to the mobile base station when they are within the radio range of the mobile base station. 

The rate of successful data delivery to the mobile base station is one of the most important performance metrics in ZNetSim. It is strongly related to the radio range of the mobile base station. The larger the radio range is, the higher rate of data collection that can be achieved. However, larger radio range also means more battery power consumption of zebra collars. Thus, achieving a high rate of data collection with a radio range that is as small as possible, is the key design issue of ZebraNet. Due to limited space, we are only interested in the rate of data delivery to the mobile base station in this work; other aspects such as bandwidth and storage issues are disregarded.

\subsection{HYPE}
HYPE is a process algebra designed to capture the behaviour of hybrid systems \cite{galpin2009hype}. A hybrid system is one in which both discrete and continuous behaviour are exhibited. In HYPE, the behaviour of a system is represented by interacting components which may consist of discrete events and continuous flows. Continuous flows are values in the system that change continuously over time whereas discrete events are actions that only take place when their activation conditions are satisfied. Once the events are activated, they can reset the value of some variables within the system. In \cite{bortolussi57hype}, stochastic event conditions are introduced into HYPE so that discrete events can be activated at a rate that is governed by an exponential distribution. We will show how this stochastic property makes HYPE suitable for patch-based modelling through the definition of the patch-based HYPE model in this paper.  

In our earlier modelling work on ZebraNet \cite{Xthesis}, we build an individual-based HYPE model which captures the dynamics of each zebra in the system. The expressiveness of HYPE can be seen from the script size of the resulting HYPE model which consists of 440 lines of definition, compared with ZNetSim which has 5941 lines of code in C. Unfortunately, the individual-based HYPE model suffers from \emph{flow and event explosion}. For example, suppose there are $n$ zebras in the system, the resulting model needs $O(n^2)$ number of continuous flows (transmission of data) and discrete events (activation of peer data exchange) to capture the data exchange between each pair of zebras. Consequently, when the number of zebras is large, it is extremely expensive to simulate the model in the simulation tool, SimHyA \cite{bortolussi2012hybrid}. This motivates us to build a patch-based model for ZebraNet which is able to model an arbitrary number of zebras.

\section{The patch-based stochastic HYPE Model}
In this section, we present the patch-based stochastic HYPE model for ZebraNet.

\subsection{Division of ZebraNet Map into Patches}
The first step is to divide the ZebraNet map into patches. As zebras get thirsty and go to their nearest water source on each day, we infer that if there was only one water source, the distance between a zebra and the water source should follow a stationary distribution. In order to validate our inference, we wrote a simulation program (in Java) with only one zebra and one water source on the map and recorded the distance between the zebra and the water source over a long period. The zebra's movement pattern in our simulation program is consistent with \cite{zebraNet}. Due to limited space, we do not introduce the zebras' movement pattern in detail in this paper.

\begin{figure} [ht]
\centering
\includegraphics[width=.6\textwidth]{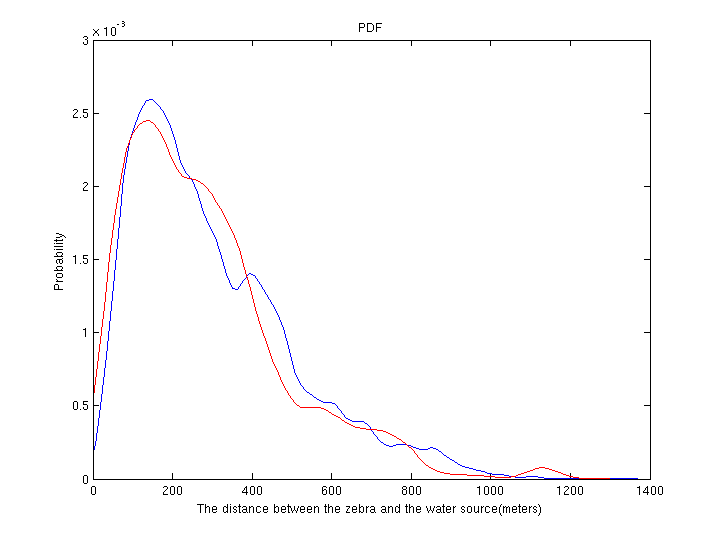}
\caption{The probability distribution of the distance between the zebra and the water source of two separate simulation runs.}
\label{fig:pdf}
\end{figure}

\begin{figure} [ht]
\centering
\includegraphics[width=.6\textwidth]{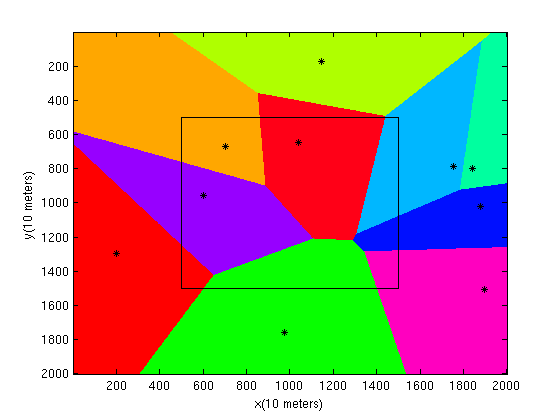}
\caption{The patch-based map of ZebraNet.}
\label{fig:mapsplit}
\end{figure}

Figure~\ref{fig:pdf} illustrates the probability distribution of the distance between the zebra and the water source in two separate simulation runs, which confirms our inference. According to the unique movement pattern of zebras, we divide the $20km\times 20km$ ZebraNet map into 10 patches based on the position of the water sources. More specifically, each point in the map belongs to the patch of its nearest water source. For instance, the map will be divided into patches as is shown in the Voronoi digram\cite{aurenhammer1991voronoi} in Figure~\ref{fig:mapsplit}. The positions of water sources tagged with stars are the Voronoi seeds; the solid black lines denote the rectangular route of the mobile base station.

%\begin{figure}[!htb]
%\minipage[t]{0.48\textwidth}
%  \includegraphics[width=\linewidth]{pdf.png}
%  \caption{The probability distribution of the distance between the zebra and the water source of two %separate simulation runs}\label{fig:pdf}
%\endminipage\hfill
%\minipage[t]{0.48\textwidth}
%  \includegraphics[width=\linewidth]{mapsplit.png}
%  \caption{The patch-based map of ZebraNet}\label{fig:mapsplit}
%\endminipage
%\end{figure}

\subsection{Key Parameters}
\label{subsection1}
Some key parameters need to be calculated for the patch-based model of ZebraNet. They are:
\begin{itemize}
\item $\alpha_{\:i}$: a zebra's contact rate with the mobile base station in the $i$th patch.
\item $\beta^{\:i}_{\:j}$: the peer contact rate between a zebra in the $i$th patch and a zebra in the $j$th patch.
\item $\gamma^{\:i}_{\:j}$: the rate at which a zebra in the $i$th patch moves to the $j$th patch.
\end{itemize} 
We wrote a Java simulation program to obtain these parameters. In the simulation, we put the 10 water sources and the mobile base station in the map. There is only one zebra placed in each patch. The movement of zebras and the mobile base station is simulated following the movement pattern described in \cite{zebraNet}. Contact events between a zebra and the mobile base station are activated when the zebra is within the radio range of the base station. Zebras' peer contact events take place when the zebras in the $i$th and $j$th patches are within the peer contact range\footnote{To make the contact events countable, the contact event (between two zebras or a zebra and the mobile base station) is prohibited for 30 minutes after a contact event occurs.}. Patch move events occur when a zebra in the $i$th patch moves towards the water source of $j$th patch. In the next day, we put this zebra migrating from the $i$th patch to the $j$th patch back to a random position in the $i$th patch again to make sure that there is always one zebra in each patch at the start of a day. As there are only 10 zebras in this simulation program, the simulation cost is quite low. More specifically, it only costs about 30 seconds to run a simulation for 10 years' simulation time length on a dual CORE i5 machine with 2GB RAM.

\subsection{Variables}
Next, we present the discrete and continuous variables in the patch-based stochastic HYPE model: \begin{itemize}
\item $N_{i}$: the current number of zebras in the $i$th patch.
\item $A_{i}$: the current age of data at the mobile base station for the $i$th patch. It indicates the length of time since the last time the mobile base station received fresh data (directly or indirectly) from the $i$th patch.
\item $A^{i}_{j}$: the current age of data at the $j$th patch for the $i$th patch. It indicates the length of time since the zebras in the $j$th patch received fresh data from the $i$th patch.
\end{itemize}
Note that within our model, we assume that each zebra shares all data in its current patch so that the age of data at the mobile base station for the $i$th patch can be captured by a single variable $A_{i}$, which is similar for $A^{i}_{j}$. This assumption is validated by a similar simple simulation program with only one water source and two zebras. The simulation result shows that there are 23,793 out of 36,500 days that these two zebras have at least one chance to contact each other. According to the flooding protocol, the frequency that one zebra gets fresh data from another zebra will increase when the number of zebras grows. As a result, it is reasonable that we treat data of zebras within one patch as an aggregation.

\subsection{Main Ingredients of the HYPE Model}
The main ingredients of a HYPE model are: \begin{inparaenum}[\itshape a\upshape)]
\item the subcomponents which consist of continuous flows and discrete events, 
\item the event conditions which presents the activation conditions for each event,
\item the controllers which indicate the constraints on events,
\item the uncontrolled system which is the combination of subcomponents, 
\item the controlled system which is constructed by synchronization of the uncontrolled system and controllers.
\end{inparaenum}
We will present these ingredients for the patch-based HYPE model for ZebraNet one by one later in this section.

\subsubsection{Subcomponents}
Continuous flows are represented by subcomponents in a HYPE model. There are two types of continuous flows in the model. The first one is the increasing of the age variables $A_{i}$, whereas the second one is the increasing of the age variables $A^{i}_{j}$. We give the definition of the subcomponents for these flows as follows:
\begin{eqnarray*}
\IncA_{i} &\rmdefl & \ev{init}\pc(A_{i},1,\const).\IncA_{i} \quad i \in (0,1,\dotsc,n)   \\
\IncA^{i}_{j} &\rmdefl & \ev{init}\pc(A^{i}_{j},1,\const).\IncA^{i}_{j} \quad i,j \in (0,1,\dotsc,n), i \neq j 
\end{eqnarray*}
where $n$ is a parameter equals to $9$, which is always the case hereinafter in this paper. A subcomponent is made up of prefixes. Each prefix consists of two actions. \emph{Events} are actions which happen either stochastically or deterministically according to their activation conditions. In the above subcomponents, the event is \ev{init}, which is the default initialisation event. \emph{Activities} are flows which influence the evolution of the continuous part of the system. An activity is defined as a tuple, $\alpha(X) = (\iota,r,I(X))$  which consists of an \emph{influence name} $\iota$, a rate of change (or \emph{influence strength}) $r$ and an \emph{influence type name} $I(X)$ which indicates how that rate is to be applied to the variable involved. In the above definition, there is only one distinct activity in each subcomponent. The influence names are the variables $A_i$ and $A^i_j$ respectively. The influence strength is $1$. The influence type name is the function $\const$\footnote{$[\const] = 1$}. This means that the age variables $A_i$ and $A^i_j$ increases constantly from the beginning of the simulation. More specifically, $\frac{\mathrm d A_i}{\mathrm d t} = 1\times 1$ and $\frac{\mathrm d A^i_j}{\mathrm d t} = 1\times 1$, where $t$ is the inherent time variable in the simulation. 

\subsection{Event Conditions}
Event conditions capture the discrete behaviour of the system. Each event condition consists of an activation condition and several variable resets. An activation condition can be either stochastic or deterministic. A deterministic activation condition is a positive boolean formula containing equalities and inequalities on system variables whereas a stochastic activation condition is a rate that is governed by an exponential distribution. A variable reset is a conjunction of equality predicates on variables $V$ and $V'$ where $V'$ denotes the new value that $V$ will have after the reset, whereas V denotes the previous value before reset.

In this model, three series of event conditions are required to capture the discrete dynamics of the system. 

First of all, we need event conditions to represent the contact events between the mobile base station and the zebras in a patch. Thus, for $0\leq i \leq n$, for Patch $i$ we have
\begin{align*}
\ec(\overline{BaseContact\_i}) = &  
(\alpha_{i}\times N_{i}, \quad A_i'=0 \wedge A_j'=min(A_{i}^{j}, A_j)\wedge \ldots) \quad j \in (0,1,\dotsc,n), j \neq i
\end{align*} 
in which the overline denotation means that the corresponding event is activated stochastically. Here, as the rate of contact between a patch and the mobile base station depends on the current number of zebras within the patch, the event $\overline{BaseContact\_i}$ will be activated at a rate of $\alpha_{i}\times N_{i}$ governed by an exponential distribution. On firing, the event will set the age of data for the $i$th patch at the base station to $0$, which means that the zebras in this patch transfer all their own data to the base station. Moreover, the age of data for other patches at the base station will also be updated if the age of data for that patch at the $i$th patch is smaller than the age of data at the base station. This captures that the zebra will transfer its stored data for other patches to the base station if it is fresher than the corresponding data at the base station.

Several event conditions are also required to represent the peer contact events between zebras in different patches. For example, for $i \in (0,1,\dotsc,n-1)$, $j \in (1,2,\dotsc,n)$ and $i < j$, the peer contact events between the zebras in Patch $i$ and Patch $j$ can be denoted by the following definition:
\begin{align*}
\ec(\overline{PeerContact\_i\_j}) = &  
(\beta^i_j \times N_i \times N_j,\quad A^{j'}_{i}=0 \wedge A_{j}^{i'}=0 \wedge A_i^{k'}=min(A_i^{k},A_j^{k}) \wedge A_j^{k'}=min(A_i^{k},A_j^{k})\\
&\wedge \ldots ) \quad k \in (0,1,\dots,n), k \neq i, k\neq j
\end{align*} 
Note that this is also a stochastic event governed by an exponential distribution. The activation rate is $\beta^i_j \times N_i \times N_j$ as it depends on the current number of zebras in both patches. Once the peer contact event between Patch $i$ and $j$ is activated, the zebras in these two patches will exchange their data. Therefore, the age of data at the $i$th patch for the $j$th patch and the age of data at the $j$th patch for the $i$th patch will be set to $0$. Meanwhile, according to the flooding protocol, the zebras will also exchange their data for other patches. Hence, the age of data for other patches at both patches will be updated to the smaller one between them.

Lastly, zebras will occasionally move to their neighbouring patches. Thus, for $i\in (0,1,\dotsc,n)$, $j \in (0,1,\dotsc,n)$ and $i \neq j$, movement of a zebra from the $i$th patch to the $j$th patch is captured by the event condition shown below:
\begin{align*}
\ec(\overline{PatchMove\_i\_j}) = &  
(\gamma^{\:i}_{\:j} \times N_i,\quad N_i'=N_i-1 \wedge N_j'=N_j+1 \wedge A_j^{i'}=0 \wedge A_j^{k'}=min(A_j^{k}, A_i^{k})\wedge \ldots)\\
&k \in (0,1,\dotsc,n), k \neq i,k \neq j
\end{align*}
$\overline{PatchMove\_i\_j}$ is also a stochastic event with activation rate $\gamma^{\:i}_{\:j} \times N_i$, as the rate is dependent of the current number of zebras in the $i$th patch. Clearly, when a zebra moves from the $i$th patch to the $j$th patch, the number of zebras in the $i$th patch will decrease by one and the number of zebras in the $j$th patch will increase by one. Moreover, the zebra will bring its data from the $i$th patch to the $j$th patch. Thus, the age of data for the $i$th patch at the $j$th patch will be set to $0$. Additionally, the age of data for other patches in the $j$th patch will also be updated if it is larger than its counterpart in the $i$th patch.

\subsection{Controllers}
Controllers are used to impose causal or temporal constraints on events in HYPE. The controllers in this model are quite straightforward. They only guarantee that the events in the model take place in parallel. We give their definition below: 
\begin{eqnarray*}
Con_{bc_{i}} & \rmdefl & \overline{BaseContact\_i}.Con_{bc_{i}} \quad i \in (0,1,...,n) \\
Con_{pc_j^i} & \rmdefl & \overline{PeerContact\_i\_j}.Con_{pc_j^i} \quad i \in (0,..,n-1), j \in (1,..,n), i<j\\
Con_{pm_j^i} & \rmdefl & \overline{PatchMove\_i\_j}.Con_{pm_j^i} \quad i \in (0,1,...,n), j \in (0,1,...,n), i\neq j\\
Con & \rmdefl &  ... Con_{bc_{i}}...||... Con_{pc_i^j}... ||  ... Con_{pm_i^j}... 
\end{eqnarray*}
where $\Con$ is the overall controller, which is the parallel combination of all the controllers in the model. 

\subsection{Controlled System}
A HYPE model consists of the uncontrolled system in cooperation with controllers. We get the uncontrolled system by synchronizing the subcomponents in the system:
\begin{eqnarray*}
\Sys & \rmdefl &  ...\IncA_{i}...\sync{\ev{init}}  ...\IncA_{i}^j... \quad i\in (0,1,..,n),j \in (0,1,..,n), i \neq j
\end{eqnarray*}
Finally, the controlled system of the model is described by:
\begin{eqnarray*}
\ZebraNetCtrl & \rmdefl & \Sys \sync{*} \ev{init}.\Con
\end{eqnarray*}

\section{Evaluation of the patch-based HYPE Model}
\label{evalu}
The evaluation of the patch-based stochastic HYPE model for ZebraNet is based on four aspects. The first aspect is the scalability of the model. One can easily find that the model size of the patch-based model will stay almost invariant as the number of zebras grows, which means that the scalability of the patch-based model is largely enhanced compared with the individual-based model. As mentioned previously, the model size is mostly decided by the number of patches in the model. According to our test, currently the simulation tool for HYPE, SimHyA can load and simulate the ZebraNet model with at most 35 patches. The second aspect is the ease of building the model. This can be seen through the model definition, in which different kinds of continuous flows and discrete events are modelled separately, and are composed to capture the dynamics of the whole system. The third aspect is the generality of the model. We believe the framework of our patch-based stochastic HYPE model can be used to model many systems with similar spatial features. For example, in the field of opportunistic networks \cite{OppNetSurvey}, systems such as DakNet \cite{pentland2004daknet}, SNC \cite{doria2002providing} and SWIM \cite{small2003shared} can also be modelled in a similar way. The last aspect is the accuracy of the model in terms of performance evaluation of the modelled system. In this model, the measure of interest is the rate of data delivery, i.e.$\:$the proportion of data collected which is successfully transferred to the mobile base station. We will explain how this measure is computed from the patch-based HYPE model later in this section. Furthermore, we will also validate the simulation result of the patch-based HYPE model by comparing it with ZNetSim.

\subsection{Rate of Data Delivery}
As mentioned above, we are interested in the rate of data collection in this model. Assume that each zebra generates $k$ amount of data every time unit, then the total amount of data generated by the zebras in the whole system is $N\times k\times t$, in which $N$ is the total number of zebras in the system, $t$ is the time length of the simulation. For Patch $i$, $A_{i}$ indicates the length of time since the last time the base station received fresh data from this patch. In other words, $A_{i}$ denotes for how long the base station has not received any fresh data from the $i$th patch. Hence, there is $N_i\times k\times A_{i}$ amount of data that has not been collected from the $i$th patch at a given moment. Therefore, the total success rate of data collection by the mobile base station is:
\begin{eqnarray*}
R=1-\frac{\sum_{i=0}^n N_i\times k \times A_i}{N\times k\times t}
\end{eqnarray*}
Clearly, the patch-based HYPE model contains enough information to compute the rate of data collection by the mobile base station. Moreover, although $A^{i}_{j}$ is not very meaningful in the context of ZebraNet, it might also be very useful in other contexts where the data delivery between patches is of interest.
   
\subsection{Simulation Result}
\begin{figure} [ht]
\centering
\includegraphics[width=.7\textwidth]{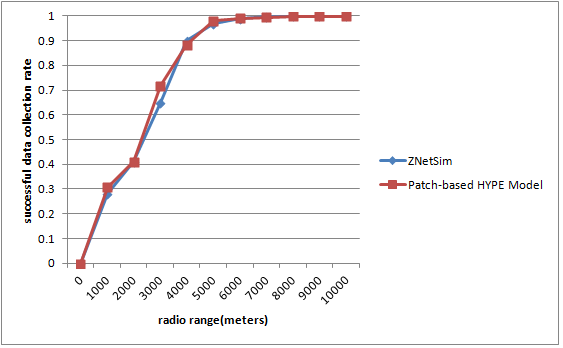}
\caption{Rate of data collection by the mobile base station}
\label{fig:succrate}
\end{figure}

The simulation tool that is used to run HYPE models is called SimHyA, which was introduced in \cite{bortolussi2012hybrid}. In our simulation, we set the time length of each simulation run to 3 months, and the number of zebras in the model to 50, to keep it consistent with ZNetSim. The radio range of the mobile base station is set between 1,000 meters and 10,000 meters in 10 steps. The parameter $\alpha_{i}$ for each radio range is obtained by the simulation program mentioned in Section \ref{subsection1}. The rate of data delivery to the mobile base station for each radio range is obtained from the simulation of the patch-based HYPE model. For each radio range, we take the average over 20 simulation runs (each simulation run costs about 30 seconds) with different random positions of water sources and random initial positions of zebras (in terms of which patch the zebras are initially in). Figure~\ref{fig:succrate} shows our simulation result compared with ZNetSim. As can be seen from the figure, our simulation result is well matched with ZNetSim. This shows the accuracy of the patch-based HYPE model for ZebraNet. 

\section{The Patch-based Mean-field Model}
\label{meanfield}
As getting performance metrics from stochastic models often requires to simulate the model a large number of times and calculate the mean, thus, it is advantageous to represent the stochastic HYPE model by an analytical model, and draw performance metrics from the analytical model which needs much smaller computation cost. In this section, we show how to use a mean-field analytical model to represent the patch-based stochastic HYPE model for ZebraNet. In the mean-field model, we capture the expected evolution of the data age for patches and the movement of zebras across patches over time as a set of ODEs. 

First of all, the value of variables can be updated both in the continuous flows (the activity tuples of subcomponents) and the discrete events (the variable resets of event conditions) in the stochastic HYPE model. Thus, by searching the appearance of a variable in all the continuous flows and discrete events (more specifically, the influence name in the activity tuples and the left side of reset equations) in the stochastic HYPE model, we can summarize how this variable evolves globally in the model. We capture this information in the \emph{evolution matrix} of the variable. 

Here, we illustrate the evolution matrices of variable $N_i$, $A_i$ and $A_j^i$ in the stochastic HYPE model for ZebraNet in Table \ref{tab:ni}, \ref{tab:ai}, \ref{tab:aji} respectively. The \emph{type of change} or \emph{influence type} indicates how the value of the variables is updated by the influence of the discrete events or continuous flows in the stochastic HYPE model respectively, whereas the \emph{rate of change} denotes the occurrence rate of the stochastic events or the influence strength of the continuous flows on the variable. If we treat the evolution of the variables caused by the discrete events in the stochastic HYPE model as continuous fluid flows, then, the evolution matrices can be used to generate ODEs that describe the evolution of variables in the stochastic HYPE model. More specifically, if we denote \emph{type of change} by $x=x+e$, the \emph{influence type} by $e$, the \emph{rate of change} by $r$, then, the influence of a discrete event or continuous flow on the variable $x$ can be denoted by $e\times r$. Therefore, the ODE to describe the evolution of variable $x$ can be obtained by summing up all the influence of the continuous flows and discrete events on $x$, which can be represented by $\frac{\mathrm d x}{\mathrm d t} = \sum e \times r$.

For example, the evolution of variables in the patch-based stochastic HYPE model for ZebraNet can be described by the following ODEs, which can be readily derived from the evolution matrices in Table \ref{tab:ni}, \ref{tab:ai}, \ref{tab:aji} respectively.
\begin{eqnarray}
\frac{\mathrm d N_i}{\mathrm d t} &=& \sum_{j\neq i} \gamma^{\:j}_{\:i} \times N_j - \sum_{j\neq i} \gamma^{\:i}_{\:j} \times N_i \quad i \in (0,1,..,n)\\
\frac{\mathrm d A_i}{\mathrm d t} &=& 1-\alpha_i\times N_i\times A_i - \sum_{j\neq i} \alpha_j\times N_j\times \frac{|A_i-A^i_j|+A_i-A^i_j}{2} \quad i \in (0,1,..,n)\\
\frac{\mathrm d A_j^i}{\mathrm d t} &=& 1-\beta^i_j\times N_i\times N_j\times A_j^i - \gamma^{\:i}_{\:j} \times N_i \times A_j^i \nonumber - \sum_{k\neq i,j} \beta_k^j \times N_j\times N_k\times \frac{|A_j^i-A_k^i|+A_j^i-A_k^i}{2} \\
& &- \sum_{k\neq i,j} \gamma^{\:k}_{\:j} \times N_k\times \frac{|A_j^i-A_k^i|+A_j^i-A_k^i}{2} \quad i \in (0,1,..,n), j \in (0,1,..,n), i \neq j
\end{eqnarray}

The ODE for $N_i$ consists of two parts. The first part captures zebras moving from other patches to Patch $i$ ($\overline{PatchMove\_j\_i}$), whereas the second part captures zebras moving from Patch $i$ to other patches ($\overline{PatchMove\_i\_j}$).

The ODE for $A_i$ consists of three terms. The first term is a constant which describes the rate of age growth over time ($\IncA_{i}$). The second term denotes the zebras in Patch $i$ directly transmitting data to the base station ($\overline{BaseContact\_i}$). The third term denotes zebras in other patches sending their data for Patch $i$ to the base station if their data for Patch $i$ is fresher than the counterpart at the base station ($\overline{BaseContact\_j}$). 

The ODE for $A_j^i$ is made up by five parts. The first part is also a constant describing the rate of age growth over time ($\IncA_j^i$). The second part denotes zebras in the $i$th patch having peer contact with zebras in Patch $j$ ($\overline{PeerContact\_i\_j}$). The third part denotes zebras moving from Patch $i$ to Patch $j$, and bringing their data to Patch $j$ ($\overline{PatchMove\_i\_j}$). The fourth part denotes zebras in the $k$th patch having peer contact with zebras in the $j$th patch ($k \neq i,k \neq j$), and the zebras in the $k$th patch have fresher data for Patch $i$, thus they transfer their fresher data to the zebras in the $j$th patch ($\overline{PeerContact\_k\_j}$). The last part denotes zebras move from the $k$th patch to the $j$th patch ($k \neq i,k \neq j$), and the zebras from the $k$th patch have fresher data for Patch $i$ than the counterpart at the $j$th patch, thus they bring their data for Patch $i$ to the $j$th patch ($\overline{PatchMove\_k\_j}$).

\begin{table}
\begin{center}
    \begin{tabular}{  c | c | c}
    Event or flow name  & Type of change or Influence type & Rate of change  \\ \hline
    $\overline{PatchMove\_i\_j}$ & $N_i=N_i-1$ & $\gamma^{\:i}_{\:j} \times N_i$ \\ \hline
    $\overline{PatchMove\_j\_i}$ & $N_i=N_i+1$ & $\gamma^j_i \times N_j$ \\
    \end{tabular}
\end{center}
\caption{The evolution matrix of $N_i$}
\label{tab:ni}
\end{table}

\begin{table}
\begin{center}
    \begin{tabular}{  c | c | c }
    Event or flow name  & Type of change or Influence type & Rate of change  \\ \hline
    $\IncA_{i}$  & $1$ & $1$   \\ \hline
    $\overline{BaseContact\_i}$ & $A_i=0$ ($A_i=A_i-A_i$) & $\alpha_{i}\times N_{i}$   \\ \hline
    $\overline{BaseContact\_j}$ & $A_i=min(A_{j}^{i}, A_i)$ ($A_i=A_i-\frac{|A_i-A^i_j|+A_i-A^i_j}{2}$) & $\alpha_{j}\times N_{j}$ \\
    \end{tabular}
\end{center}
\caption{The evolution matrix of $A_i$}
\label{tab:ai}
\end{table}

\begin{table}
\begin{center}
    \begin{tabular}{  c | c | c }
    Event or flow name & Type of change or Influence type & Rate of change  \\ \hline
    $\IncA_j^i$  & $1$ & $1$   \\ \hline
    $\overline{PeerContact\_i\_j}$ & $A_j^i=0$ ($A_j^i=A_j^i-A_j^i$) & $\beta^i_j \times N_i \times N_j$   \\ \hline
    $\overline{PatchMove\_i\_j}$ & $A_j^i=0$ ($A_j^i=A_j^i-A_j^i$) & $\gamma^{\:i}_{\:j} \times N_i$ \\ \hline
    $\overline{PeerContact\_k\_j}$ & $A_j^{i}=min(A_k^{i},A_j^{i})$ ($A_j^i=A_j^i-\frac{|A_j^i-A_k^i|+A_j^i-A_k^i}{2}$) & $ \beta_k^j \times N_j\times N_k$   \\ \hline
    $\overline{PatchMove\_k\_j}$ & $A_j^{i'}=min(A_j^{i}, A_k^{i})$ ($A_j^i=A_j^i- \frac{|A_j^i-A_k^i|+A_j^i-A_k^i}{2}$) & $ \gamma^{\:k}_{\:j} \times N_k$ \\
    \end{tabular}
\end{center}
\caption{The evolution matrix of $A_j^i$}
\label{tab:aji}
\end{table}

\subsection{Analysis of the Mean-Field Model}
\begin{figure} [!htp]
\begin{center}
\mbox{
\subfigure[1 simulation run compared with mean-field analysis.]{
\includegraphics[width=.46\textwidth]{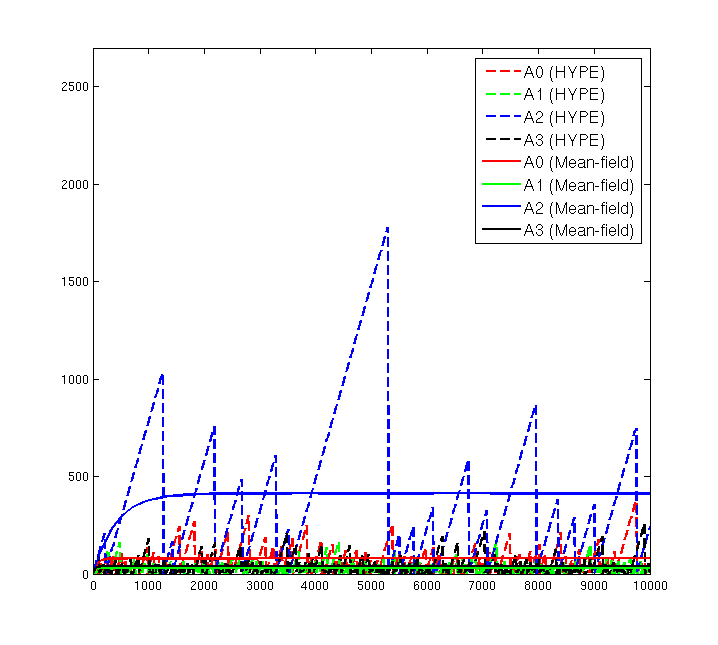}
\label{fig:hypeada}
}\quad

\subfigure[average value of 10 simulation runs compared with mean-field analysis.]{
\includegraphics[width=.46\textwidth]{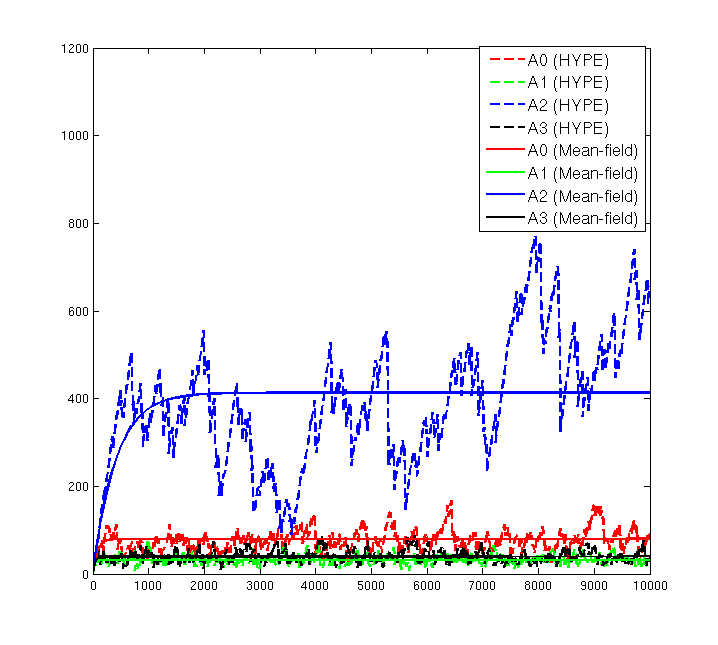}
\label{fig:odeada}
}} \\
\mbox{
\subfigure[average value of 50 simulation runs compared with mean-field analysis.]{
\includegraphics[width=.46\textwidth]{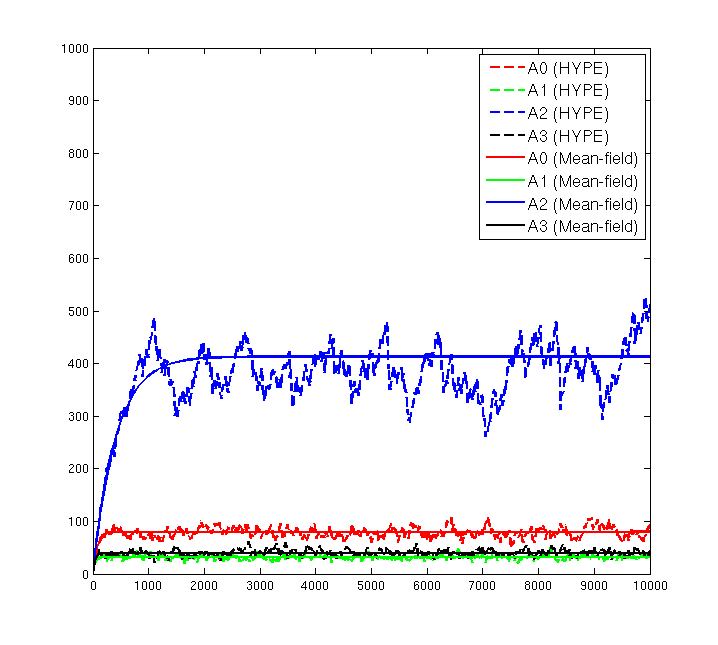}
\label{fig:50runs}
}\quad

\subfigure[average value of 100 simulation runs compared with mean-field analysis.]{
\includegraphics[width=.46\textwidth]{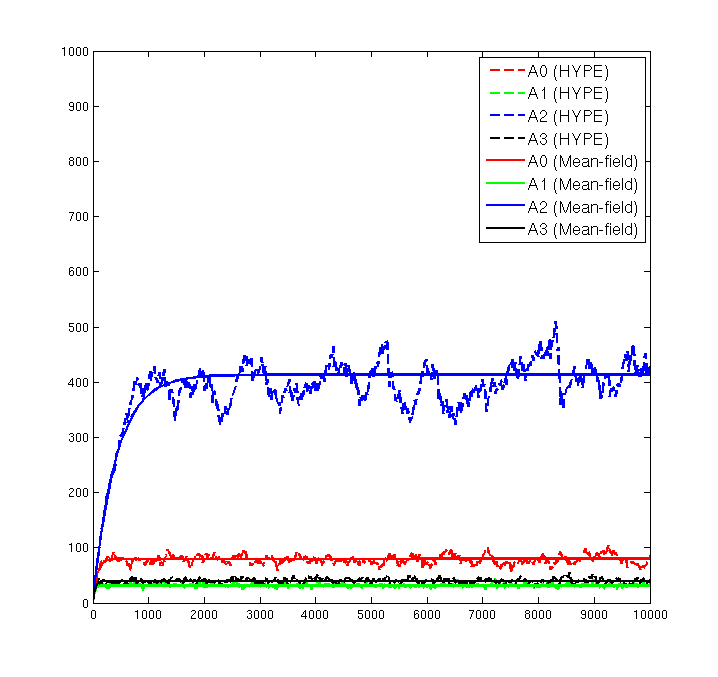}
\label{fig:100runs}
}}
\caption{The trajectory of the average value of age of data for Patch 0,1,2,3 at the mobile base station over multiple number of simulation runs of the patch-based stochastic HYPE model compared with the mean-field analysis.}
\label{fig:conSB}
\end{center}
\end{figure}

The mean-field model is the fluid flow approximation of the stochastic HYPE model. It treats the evolution of the variables caused by the discrete events in the stochastic HYPE model as fluid flows. As a result, the mean-field model gives the expected average value of the variables in the stochastic HYPE model over an infinite number of simulation runs. Figure~\ref{fig:conSB} compares the trajectories of the average value of $A_0$, $A_1$, $A_2$, $A_3$ generated by the stochastic HYPE model over different numbers of simulation runs with the corresponding trajectories generated by the mean-field model. It is clear that with more simulation runs, the trajectories generated by the stochastic HYPE model become closer to the trajectories from the mean-field model. Thus, the mean-field model provides an efficient approach to analyse the expected average behaviour of the modelled system.

\section{Related Works}
\label{relatedworks}
There has been some previous work which adopted a similar approach to model spatially distributed systems. For instance, in \cite{paof:06}, the authors use the modelling framework asCSL \cite{baier2004model} to characterize delay in a generic Delay-Tolerant network comprising a mixture of fixed and mobile nodes, in which they also aggregate nodes as patches (islands). In \cite{chaintreau2009age}, a patch-based mean-field model is developed for a gossip network system in which data aging using real data collected from cabs in the San Francisco Bay area is studied. The region is divided into 15 regular patches in a grid with a sixteenth patch representing the rest of the world. In this work, the authors also use the age of data to measure the efficiency of information dissemination in the network system. Our work is distinguished from the previous works by the modelling language we used to model the system, the process algebra stochastic HYPE. By using stochastic HYPE, the patch-based model of the complex spatially distributed system can be constructed intuitively and compositionally, which significantly reduces the burden on the modeller. 

The mean-field model, which is the fluid flow approximation of the stochastic HYPE model, is inspired by the work in \cite{hillston2005fluid}, where the author presented a systematic approach to translate discrete PEPA models to fluid flows described by ODEs. The modelling framework, Markovian Agents (MAs), which has been widely used to model spatially distributed wireless sensor networks \cite{gribaudo2008analysis, bruneo2012markovian}, is also able to derive patch-based mean-field analytical model for the spatially distributed systems. In the MA formalism, each agent is described by a CTMC. Agents have their location attributes and can interact with other agents both locally and in other locations by message passing. The ODEs are derived to describe the evolution of the agent populations in the same states and locations, in which fluid analysis method is also used. More recently, a process algebra, MASSPA, has been defined for this formalism in \cite{guenther2011higher}. In this work, we adopt a similar approach to derive fluid analysis for a hybrid model in which not only discrete behaviour, but also continuous behaviour is captured.

\section{Conclusion and Future Work}
\label{conclusions}
We have shown how to use the process algebra, stochastic HYPE to build a patch-based hybrid model for a spatially distributed system, ZebraNet. The merit of the model is that it significantly improves the scalability of the model but without significant loss of accuracy, compared with the individual-based simulation model, ZNetSim. The expressiveness of stochastic HYPE for patch-based hybrid modelling can be seen through the definition of the model, in which various continuous flows and discrete events are defined separately and are easily composed to capture the dynamics of the whole system. 

Additionally, we use the evolution matrices of the variables to derive a mean-field analytical model to represent the patch-based stochastic HYPE model for ZebraNet. The mean-field model is the fluid flow approximation of the stochastic HYPE model. It can give efficient analysis of the expected average behaviour of the modelled system over infinite simulation runs. As the mean-field model is efficient yet accurate in many cases, it is advantageous to design approaches to derive mean-field models from the definition of stochastic HYPE models systematically. We will aim to formally define this derivation and add it as a new feature of the SimHyA tool for analysing stochastic HYPE models in the near future.

\section{Acknowledgments}
The author would like to thank Jane Hillston, Vashti Galpin, Stephen Gilmore and  Luca Bortolussi for their helpful comments on an earlier draft of this work. This work is supported by the EU project QUANTICOL, 600708.

\bibliographystyle{eptcs}
\bibliography{patchbased}
\end{document}